\documentclass[english,12pt]{article}
\usepackage[cp1251]{inputenc}
\usepackage{babel}
\usepackage{amsfonts, amsmath}

\newcommand{\be}{\begin{equation}} \newcommand{\ee}{\end{equation}}

\newcommand{\gsim}{\mathrel{\hbox{\rlap{\lower.55ex\hbox{$\sim$}} \kern-.3em \raise.4ex \hbox{$>$}}}}

\begin{document}
\begin{center}
{\bf Minimal Length and the Existence of Some Infinitesimal Quantities in Quantum Theory and Gravity}\\
\vspace{5mm} A.E.Shalyt-Margolin \footnote{E-mail:
a.shalyt@mail.ru; alexm@hep.by}\\ \vspace{5mm} \textit{National
Centre of Particles and High Energy Physics, Pervomayskaya Str.
18, Minsk 220088, Belarus}
\end{center}
 \rm\normalsize \vspace{0.5cm}
\begin{abstract}
In this work it is demonstrated that, provided a theory involves a
minimal length, this theory must be free from such infinitesimal
quantities as infinitely small variations in surface of the
holographic screen, its volume, and entropy. The corresponding
infinitesimal quantities in this case must be replaced by the
«minimal variations possible» -- finite quantities dependent on
the existent energies. As a result, the initial low-energy theory
(quantum theory or general relativity) inevitably must be replaced
by a minimal-length theory that gives very close results but
operates with absolutely other mathematical apparatus.
\end{abstract}

At the present time all high-energy generalizations (limits) of
the basic «components» in fundamental physics (quantum theory
\cite{Peskin} and gravity \cite{Einst1}) of necessity lead to a
minimal length on the order of the Planck length $l_{min}\propto
l_P$. This follows from a string theory \cite{Ven1}--
\cite{Polch}, loop quantum gravity \cite{Rovel}, and other
approaches \cite{QG1}--\cite{Kempf}.
\\ But it is clear that, provided a minimal length exists,
it is existent at all the energy scales and not at high (Planck’s)
scales only.
\\ What is inferred on this basis for real physics? At least,
it is suggested that the use of infinitesimal quantities
$dx_{\mu}$ in a mathematical apparatus of both quantum theory  and
gravity is incorrect, despite the fact that both these theories
give the results correlating well with the experiment (for
example, \cite{Hawk-Penr1}).
\\ Indeed, in all cases the infinitesimal quantities $dx_{\mu}$
bring about an infinitely small length $ds$ \cite{Einst1}
\begin{equation}\label{Introd 1}
ds^{2}=g_{\mu\nu}dx_{\mu}dx_{\nu}
\end{equation}
that is inexistent because of $l_{min}$.
\\ The same is true for any function $\Upsilon$ dependent only on
{\bf } different parameters $L_{i}$ whose dimensions of length of the exponents
are equal to or greater than 1 $\nu_{i}\geq 1$
\begin{equation}\label{Introd 2}
\Upsilon \equiv \Upsilon(L^{\nu_{i}}_{i}).
\end{equation}
Obviously, the infinitely small variation $d\Upsilon$ of
$\Upsilon$ is senseless as, according to (\ref{Introd 2}), we have
\begin{equation}\label{Introd 2.1}
d\Upsilon \equiv d\Upsilon(\nu_{i}L^{\nu_{i}-1}_{i}dL_{i}).
\end{equation}
But, because of $l_{min}$, the infinitesimal quantities
$dL_{i}$ make no sense and hence $d\Upsilon$ makes no sense too.
\\ Instead of these infinitesimal quantities it seems reasonable to denote them as
«minimal variations possible» $\Delta_{min}$ of the quantity $L$
having the dimension of length,  i.e. the quantity
\begin{equation}\label{Introd 2.1new}
\Delta_{min}L=l_{min}.
\end{equation}
And then \begin{equation}\label{Introd 2.2}
\Delta_{min}\Upsilon \equiv
\Delta_{min}\Upsilon(\nu_{i}L^{\nu_{i}-1}_{i}\Delta_{min}L_{i})=\Delta_{min}\Upsilon(\nu_{i}L^{\nu_{i}-1}_{i}l_{min}).
\end{equation}
However, the «minimal variations possible» of any quantity having
the dimensions of length (\ref{Introd 2.1new}) which are equal to
$l_{min}\propto l_P$ require, according to the Heisenberg
Uncertainty Principle (HUP) \cite{Heis1}, maximal momentum
$p_{max}\propto P_{Pl}$ and energy $E_{max}\propto E_P$. Here
$l_P,P_{Pl},E_P$ -- Planck’s length, momentum, and energy,
respectively.
\\ But at low energies (far from the Planck energy) there are
no such quantities and hence in essence
$\Delta_{min}L=l_{min}\propto l_P$ (\ref{Introd 2.1new})
corresponds to the high-energy (Planck’s) case only.
\\ For the energies lower than Planck’s energy,
the «minimal variations possible» $\Delta_{min}L$ of the quantity
$L$ having the dimensions of length must be greater than $l_{min}$
and dependent on  the present $E$
\begin{equation}\label{Introd 2.3}
\Delta_{min}\equiv\Delta_{min,E},\Delta_{min,E}L >l_{min}.
\end{equation}
Besides, as we have a minimal length unit $l_{min}$, it is clear
that any quantity having the dimensions of length is «quantized»,
i.e. its value measured in the units $l_{min}$ equals an integer
number and we have
\begin{equation}\label{Introd 2.4}
L=N_{L}l_{min},
\end{equation}
where $N_{L}$-- positive integer number.
\\ The problem is, how the «minimal variations possible»
$\Delta_{min,E}$ (\ref{Introd 2.3}) are dependent on the energy
or, what is the same, on the scales of the measured lengths?
\\To solve the above-mentioned problem, {\bf initially} we
can use the Space-Time Quantum Fluctuations (STQF) with regard to
quantum theory and gravity.
\\The definition (STQF) is closely
associated with the notion of «space-time foam». The notion
«space-time foam», introduced by J. A. Wheeler about 60 years ago
for the description and   investigation of physics at Planck’s
scales (Early Universe) \cite{Wheel1},\cite{mis73}, is fairly
settled. Despite the fact that in the last decade numerous works
have been devoted to physics at Planck’s scales within the scope
of this notion, for example \cite{Gar1}--\cite{dio89}, by this
time still their no clear understanding of the «space-time foam»
as it is.
\\ On the other hand, it is undoubtful that a quantum theory
of the  Early Universe should be a deformation of the well-known
quantum theory.
\\ In my works with the colleagues \cite{shalyt1}--\cite{shalyt9}
I has put forward one of the possible approaches to resolution of
a quantum theory at Planck’s scales on the basis of the density
matrix deformation.
\\In accordance with the modern concepts, the
space-time foam \cite{mis73} notion forms the basis for space-time
at Planck’s scales (Big Bang). This object is associated with the
quantum fluctuations generated by uncertainties in measurements of
the fundamental quantities, inducing uncertainties in any distance
measurement. A precise description of the space-time foam is still
lacking along with an adequate quantum gravity theory. But for the
description of quantum fluctuations we have a number of
interesting methods (for example,
\cite{wigner},\cite{found}--\cite{dio89}).
\\In what follows, we use the terms and symbols from \cite{Ng3}.
Then for the fluctuations $\widetilde{\delta} l$ of the distance
$l$ we have the following estimate:
\begin{equation}\label{Fluct 1}
(\widetilde {\delta} l)_{\gamma} \gsim l_P^{\gamma}
l^{1-\gamma}=l_P (\frac{l}{l_P})^{1-\gamma}=l
(\frac{l_P}{l})^{\gamma}=l \lambda_{l}^{\gamma},
\end{equation}
or  that  same  one
\begin{equation}\label{Fluct 1.new}
|(\widetilde {\delta} l)_{\gamma}|_{min}=\beta l_P^{\gamma}
l^{1-\gamma}=\beta l_P (\frac{l}{l_P})^{1-\gamma}=\beta l
\lambda_{l}^{\gamma},
\end{equation}
where $0<\gamma\leq 1$, coefficient $\beta$ is  of  order 1 and
$\lambda_{l}\equiv l_P/l$.
\\From  (\ref{Fluct 1}),(\ref{Fluct 1.new}),
we can derive the quantum fluctuations for all the primary
characteristics, specifically for the time $(\widetilde{\delta}
t)_{\gamma}$, energy $(\widetilde{\delta} E)_{\gamma}$, and
metrics $(\widetilde{\delta} g_{\mu\nu})_{\gamma}$. In particular,
for $(\widetilde{\delta} g_{\mu\nu})_{\gamma}$ we can use formula
(10) in \cite{Ng3}
\begin{equation}\label{Fluct 1.4}
(\widetilde{\delta} g_{\mu\nu})_{\gamma} \gsim \lambda^{\gamma}.
\end{equation}
Further in the text is assumed that the theory involves a minimal
length on the order of Planck’s length
\\
$$l_{min} \propto l_P$$
or that is the same
\begin{equation}\label{Min1}
l_{min}=\xi l_P,
\end{equation}
where the coefficient $\xi$ is  on the order of unity too.
\\In this case the origin of the minimal length is not important.
For simplicity, we assume that it comes from the Generalized
Uncertainty Principle (GUP) that is an extension of HUP for
Planck’s energies, where gravity must be taken into consideration
\cite{Ven1}--\cite{Kempf}:
\begin{equation}\label{GUP1}
\triangle x\geq\frac{\hbar}{\triangle p}+\alpha^{\prime}
l_{P}^2\frac{\triangle p}{\hbar}.
\end{equation}
Here $\alpha^{\prime}$ is the model-dependent dimensionless
numerical factor.
\\Inequality (\ref{GUP1}) leads to the minimal length $l_{min}= \xi l_P=2\surd \alpha^{\prime}l_P$.
\\Therefore, in this case replacement of Planck’s length by the minimal
length in all the above formulae is absolutely correct and is used
without detriment to the generality \cite{Fluct1}
\begin{equation}\label{GUP1.new}
l_P \rightarrow l_{min}.
\end{equation}
Thus, $\lambda_{l}\equiv l_{min}/l$  and then (\ref{Fluct 1})--
(\ref{Fluct 1.4}) upon the replacement of (\ref{GUP1.new}) remain
unchanged.
\\ As noted in the overview \cite{Ng3}, the value $\gamma=2/3$
derived in \cite{wigner1,wigner} is totally consistent with the
Holographic Principle \cite{Hooft1}--\cite{Bou1}.
\\The following points of importance should be noted \cite{Fluct1}:
\\1.1)It is clear that {\bf at Planck’s scales, i.e. at the minimal
length scales}
\begin{equation}\label{Planck 1.1}
l \rightarrow l_{min}
\end{equation}
models  for different values of the parameter $\gamma$  are
coincident.
\\
\\1.2) As noted, specifically in (\ref{Introd 2.4}),
{\bf provided some quantity has a minimal measuring unit, values
of this quantity are multiples of this unit}.
\\Naturally, any quantity having a minimal measuring unit is uniformly discrete.
\\The latter property is not met, in particular, by the energy $E$.
\\As $E\sim 1/l$, where $l$ -- measurable scale, {\bf the energy $E$
is a discrete  but  nonuniform quantity}. It is clear that the
difference between the adjacent values of   $E$
 is the less the lower $E$.  In other words, for $l\gg l_{min}$ i.
 e.
\begin{equation}\label{Fluct1.5}
E\ll E_P
\end{equation}
$E$ becomes a practically continuous quantity.
\\
\\1.3) In fact, the parameter $\lambda_{l}$ was introduced
earlier in papers  \cite{shalyt1}--\cite{shalyt9} as a deformation
parameter on going from the canonical quantum mechanics to  the
quantum mechanics at Planck’s scales (early Universe) that is
considered to be the quantum mechanics  with the fundamental
length (QMFL):
\begin{equation}\label{D1}
0<\alpha_{x}=l_{min}^{2}/x^{2}\leq 1/4,
\end{equation}
where $x$ is the measuring scale, $l_{min}\sim l_{p}$.
\\{\bf The deformation is understood as an extension of a particular theory
by inclusion of one or several additional parameters in such a way
that the initial theory appears in the limiting transition}
\cite{Fadd}.
\\ Obviously, everywhere, apart from the limiting point $\lambda_{x}=1$
or $x=l_{min}$, we have
\begin{equation}\label{D1.1}
\lambda_{x}=\sqrt{\alpha_{x}},
\end{equation}
From (\ref{D1}) it is seen that at the limiting point $x=l_{min}$
the parameter $\alpha_{x}$  is not defined due to the appearance
of singularity \cite{shalyt1}--\cite{shalyt9}. But at this point
its definition may be extended (regularized).
\\ The parameter $\alpha_{l}$ has the following clear physical meaning:
\begin{equation}\label{Beken1}
\alpha^{-1}_{l}\sim S^{BH},
\end{equation}
where
\begin{equation}\label{Beken2}
S^{BH}=\frac{A}{4l^{2}_{p}}
\end{equation}
is the well-known Bekenstein-Hawking formula for the  black hole
entropy in the semiclassical approximation
\cite{Bek1},\cite{Hawk1} for the black-hole  event horizon surface
$A$, with the characteristics linear dimension («radius») $R=l$.
This is especially obvious in the spherically-symmetric case.
\\ In what follows we use both parameters: $\lambda_{x}$  and  $\alpha_{x}$.
\\ Turning back to the introductory section of this work and to the definition
$\Delta_{min,E}L$, we assume the following:
\begin{equation}\label{D1.1new}
|\Delta_{min,E}L|=|(\widetilde {\delta} L)_{\gamma}|_{min},
\end{equation}
where $|(\widetilde {\delta} L)_{\gamma}|_{min}$ is from formula
(\ref{Fluct 1.new}), $\gamma$ -- fixed parameter from formulae (\ref{Fluct 1}),
(\ref{Fluct 1.new}), and   $E=c\hbar/L$.
\\ In physics, and in thermodynamics in particular,
the \textit{extensive quantities or parameters} are those
proportional to the mass of a system or to its volume. Proceeding
from the definition (\ref{Introd 2}) of the function
$\Upsilon(L^{\nu_{i}}_{i})$, one can generalize this notion,
taking as a {\bf Generalized Extensive Quantity (GEQ)} of some
spatial system $\Omega$ the function dependent only on the linear
dimensions of this system, with the exponents no less than 1.
\\ The function $\Upsilon(L^{\nu_{i}}_{i}),\nu_{i}\geq 1$  (\ref{Introd 2})
is {\bf GEQ}  of the system $\Omega$ with the characteristic
linear dimensions $L_{i};i=1,..,n$ or, identically, a sum of the
systems $\Omega_{i};i=1,..,n$, each of which has its individual
characteristic linear dimension $L_{i}$.
\\ Then from the initial formulae ({\ref{Introd 2})--({\ref{Introd
2.3}) it directly follows that, provided the minimal length $l_{min}$
 is existent, {\bf there are no} infinitesimal variations of GEQ.
\\ In the first place, this is true for such simplest objects
as the $n$-dimensional sphere $B_{n},n\geq 2$, whose surface area
(area of the corresponding hypersphere $S_{n}$) and volume $V_{n}$
represent GEQs and are equal to the following:
\begin{equation}\label{D3}
S_{n}=nC_{n}R^{n-1};V_{n}=C_{n}R^{n},
\end{equation}
where $R$ -- radius of a sphere the length of which is a
characteristic linear size,
$C_{n}=\pi^{n/2}/\Gamma(\frac{n}{2}+1)$ , and $\Gamma(x)$ is a
gamma-function.
\\ Of course, the same is true for   the $n$-dimensional cube (or hypercube)
$A_{n}$; its surface area and its volume are GEQs, and a length of
its edge is a characteristic linear dimension.
\\Provided  $l_{min}$  exists, there are no infinitesimal
increments for both the surface area and volume of $A_{n}$ or
$B_{n}$; only minimal variations possible for these quantities are
the case.
\\ In what follows we consider only the spatial systems whose
surface areas and volumes are GEQs.
\\
\\ Let us consider a simple but very important example of gravity in horizon spaces.
\\Gravity and thermodynamics of horizon spaces and their interrelations
are currently most actively studied
\cite{Padm}--\cite{Padman2009}. Let us consider a relatively
simple illustration – the case of a   static spherically-symmetric
horizon in space-time, the  horizon being described by the metric
\begin{equation}\label{GT9}
ds^2 = -f(r) c^2 dt^2 + f^{-1}(r) dr^2 + r^2 d\Omega^2.
\end{equation}
The horizon location will be given by a simple zero of the
function $f(r)$, at the radius  $r=a$.
\\This case is studied in detail by T.Padmanabhan in his works
\cite{Padm,Padm13} and by the author of this paper in
\cite{shalyt-IJMPD}. We use the notation system of \cite{Padm13}.
Let, for simplicity, the space be denoted as $\cal H$.
\\It is known that for horizon spaces one can introduce
the temperature that can be identified with an analytic
continuation to imaginary time. In the case under consideration
(\cite{Padm13}, eq.(116))
\begin{equation}\label{GT10}
k_BT=\frac{\hbar cf'(a)}{4\pi}.
\end{equation}
Therewith, the condition $f(a)=0$ and $f'(a)\ne 0$ must be
fulfilled.
\\ Then at the horizon $r=a$ Einstein's equations have the  form
\begin{equation}\label{GT11}
\frac{c^4}{G}\left[\frac{1}{ 2} f'(a)a - \frac{1}{2}\right] = 4\pi
P a^2
\end{equation}
where $P = T^{r}_{r}$ is the trace of the momentum-energy tensor
and radial pressure.
\\ Now we proceed to the variables «${\alpha}$» from the formula
 (\ref{D1}) to consider (\ref{GT11}) in a new notation, expressing $a$
in terms of the corresponding deformation parameter $\alpha$. In
what follows we omit the subscript in formula (\ref{D1}) of
$\alpha_{x}$, where the context implies which index is the case.
In particular, here we use $\alpha$ instead of  $\alpha_{a}$. Then
we have
\begin{equation}\label{GT13}
a=l_{min}\alpha^{-1/2}.
\end{equation}
Therefore,
\begin{equation}\label{GT14}
f'(a)=-2l^{-1}_{min}\alpha^{3/2}f'(\alpha).
\end{equation}
Substituting this into (\ref{GT11}) we obtain in the considered
case of Einstein's equations in the «${\alpha}$--representation»
the following \cite{shalyt-IJMPD}:
\begin{equation}\label{GT16}
\frac{c^{4}}{G}(-\alpha f'(\alpha)-\frac{1}{2})=4\pi
P\alpha^{-1}l^{2}_{min}.
\end{equation}
Multiplying the left- and right-hand sides of the last equation by
$\alpha$, we get
\begin{equation}\label{GT16.1}
\frac{c^{4}}{G}(-f'(\alpha)\alpha^{2}-\frac{1}{2}\alpha)=4\pi
Pl^{2}_{min}.
\end{equation}
L.h.s. of (\ref{GT16.1}) is dependent on $\alpha$. Because of
this, r.h.s. of (\ref{GT16.1}) must be dependent on $\alpha$ as
well, i. e. $P=P(\alpha)$, i.e
\begin{equation}\label{GT16.1new}
\frac{c^{4}}{G}(-f'(\alpha)\alpha^{2}-\frac{1}{2}\alpha)=4\pi
P(\alpha)l^{2}_{min}.
\end{equation}
Note that in this specific case the parameter $\alpha$ within
constant factors is coincident with the Gaussian curvature $K_{a}$
\cite{Geometry} corresponding to $a$:
\begin{equation}\label{GT16.C}
 \frac{l^{2}_{min}}{a^{2}}=l^{2}_{min}K_{a}.
\end{equation}
Substituting r.h.s of (\ref{GT16.C}) into (\ref{GT16.1new}), we
obtain the Einstein equation on horizon, in this case in terms of
the   Gaussian curvature
\begin{equation}\label{GT16.2}
\frac{c^{4}}{G}(-f'(K_{a})K_{a}^{2}-\frac{1}{2}K_{a})=4\pi
P(K_{a}).
\end{equation}
This means that up to the constants
\begin{equation}\label{GT16.2new}
-f'(K_{a})K_{a}^{2}-\frac{1}{2}K_{a}= P(K_{a}),
\end{equation}
i.e. the Gaussian curvature  $K_{a}$ is a solution of Einstein
equations  in this case. Then we examine different cases of the
solution (\ref{GT16.2new}) with due regard for considerations of
formula (\ref{D1.1new}).
\\
\\2.1) First, let us assume that $a\gg l_{min}$. As, according (\ref{Introd 2.4}),
the radius $a$ is quantized, we have $a=N_{a}l_{min}$ with the
natural number $N_{a}\gg 1$. Then it is clear that the Gaussian
curvature $K_{a}=1/a^{2}\approx 0$ takes a (nonuniform) discrete
series of values close to zero, and, within the factor
$1/l^{2}_{min}$, this series represents inverse squares of natural
numbers
\begin{equation}\label{GT16.2ser}
(K_{a})= ({\frac{1}{N^{2}_{a}},\frac{1}{(N_{a}\pm
1)^{2}},\frac{1}{(N_{a}\pm 2)^{2}},...}).
\end{equation}
Let us return to formulas (\ref{Fluct 1.new}),(\ref{D1.1new})  for
$l=a$
\begin{equation}\label{Fluct 1.new-a}
|((\widetilde {\delta} a)_{\gamma})_{min}| =  \beta N_{a}l_{min}
N_{a}^{-\gamma}= \beta N_{a}^{1-\gamma}l_{min},
\end{equation}
where $\beta$ in this case contains the proportionality factor
that relates $l_{min}$ and  $l_{P}$.
\\ Then, according to (\ref{D1.1new}), $a_{\pm1}$ is a measurable value
 of the radius  $r$  following after $a$,  and we have
\begin{equation}\label{Fluct 1.new-a1}
(a_{\pm1})_{\gamma}\equiv a\pm ((\widetilde {\delta}
a)_{\gamma})_{min} = a \pm \beta
N_{a}^{1-\gamma}l_{min}=N_{a}(1\pm \beta N_{a}^{-\gamma})l_{min}.
\end{equation}
But, as $N_{a}\gg 1$, for sufficiently large $N_{a}$ and fixed
$\gamma$, the bracketed expression in  r.h.s. (\ref{Fluct
1.new-a1}) is close to 1:
\begin{equation}\label{Fluct 1.new-a2}
1\pm \beta N_{a}^{-\gamma}\approx 1.
\end{equation}
Obviously, we get
\begin{equation}\label{Fluct 1.new-a3}
\lim\limits_{N_{a}\rightarrow\infty} (1\pm \beta
N_{a}^{-\gamma})\rightarrow 1.
\end{equation}
As a result, the Gaussian curvature $K_{a_{\pm1}}$ corresponding
to $r=a_{\pm1}$
\begin{equation}\label{Fluct 1.new-a4}
K_{a_{\pm1}}=1/a_{\pm1}^{2}\propto\frac{1}{N_{a}^{2}(1\pm \beta
N_{a}^{-\gamma})^{2}}
\end{equation}
in the case under study is only slightly different from $K_{a}$.
\\And this is the case for sufficiently large values of $N_{a}$,
for any value of the parameter $\gamma$ , for $\gamma=1$ as well,
corresponding to the absolute minimum of fluctuations $\approx
l_{min}$,or more precisely -- to $\beta l_{min}$. However, as all
the quantities of the length dimension are quantized and the
factor $\beta$ is on the order of 1, actually we have $\beta=1$.
\\ Because of this, provided the minimal length is involved,
$l_{min}$ (\ref{Fluct 1.new})  is read as
\begin{equation}\label{Fluct 1.new2} |(\widetilde {\delta}
l)_{1}|_{min}=l_{min}.
\end{equation}
But, according to (\ref{Min1}), $l_{min}=\xi l_P$ is on the order
of Planck’s length, and it is clear that the fluctuation
$|(\widetilde {\delta} l)_{1}|_{min}$ corresponds to Planck’s
energies and Planck’s scales. The Gaussian curvature $K_{a}$, due
to its smallness ($K_{a}\ll 1$ up to the constant factor
$l_{min}^{-2}$) and smooth variations independent of  $\gamma$
(formulas (\ref{Fluct 1.new-a1})--(\ref{Fluct 1.new-a4})), is {\bf
insensitive} to the differences between various values of
$\gamma$.
\\Consequently, for sufficiently small Gaussian curvature $K_{a}$
we can take any parameter  from the interval $0<\gamma\leq 1$ as
$\gamma$.
\\ It is obvious that the case $\gamma = 1$,
i. e. $|(\widetilde {\delta} l)_{1}|_{min}=l_{min}$, is associated
with infinitely small variations $da$  of the radius $r=a$  in the
Riemannian geometry.
\\ Since then   $K_{a}$  is varying practically continuously, in terms of $K_{a}$
up to the constant factor we can obtain the following expression:
\begin{equation}\label{GT12.1}
d[L(K_{a})]= d[P(K_{a})],
\end{equation}
Where have
\begin{equation}\label{GT12.2}
L(K_{a})=-f'(K_{a})K_{a}^{2}-\frac{1}{2}K_{a},
\end{equation}
i. e.  l.h.s of (\ref{GT16.2}) (or (\ref{GT16.2new})).
\\But in fact, as in this case the energies are low,
it is more correct to consider
\begin{equation}\label{GT12.3}
L((K_{a_{\pm1}})_{\gamma})-L(K_{a})=
[P(K_{a_{\pm1}})_{\gamma}]-[P(K_{a})]\equiv F_{\gamma}[P(K_{a})],
\end{equation}
where $\gamma<1$,rather than (\ref{GT12.1}).
\\ In view of the foregoing arguments 2.1), the difference between
 (\ref{GT12.3}) and (\ref{GT12.1}) is insignificant and it is perfectly
 correct to use   (\ref{GT12.1}) instead of (\ref{GT12.3}).
\\2.2) Now we consider the opposite case or the transition to  the
{\bf ultraviolet limit}
\begin{equation}\label{GT12.3.4}
a \rightarrow l_{min}=\kappa l_{min},
\end{equation}
i.e.
 \begin{equation}\label{GT12.3.4new}
 a=\kappa l_{min}.
 \end{equation}
Here $\kappa$ is on the order of 1.
\\Taking into consideration point 1.1)
stating that in this case models for different values of the
parameter $\gamma$ are coincident, by formula (\ref{Fluct 1.new2})
 for any $\gamma$ we have
\begin{equation}\label{GT12.3.5}
|(\widetilde {\delta} l)_{\gamma}|_{min}=|(\widetilde {\delta}
l)_{1}|_{min}=l_{min}.
\end{equation}
But in this case the Gaussian curvature $K_{a}$  is not a «small
value» continuously dependent on  $a$ , taking, according to
(\ref{Fluct 1.new-a4}), a discrete series of values
$K_{a},K_{a_{\pm1}},K_{a_{\pm2}},..$.
\\ Yet (\ref{GT11}), similar to   (\ref{GT16.2})
((\ref{GT16.2new})), is valid   in the semiclassical approximation
only, i.e. at  {\bf low energies}.
\\Then in accordance with the above arguments,
the limiting transition to {\bf high energies} (\ref{GT12.3.4})
 gives a discrete chain of equations or a single equation with a
discrete set of solutions as follows:
\\
$$-f'(K_{a})K_{a}^{2}-\frac{1}{2}K_{a}= \Theta(K_{a});$$
\\
$$-f'(K_{a\pm1})K^{2}_{a\pm1}-\frac{1}{2}K_{a\pm1}=
\Theta(K_{a\pm1});$$
\\and so on. Here $\Theta(K_{a})$ -- some function that
in the limiting transition to low energies must reproduce the
low-energy result to a high degree of accuracy, i.e. $P(K_{a})$
appears for $a\gg l_{min}$ from formula (\ref{GT16.2new})
\begin{equation}\label{GT12.3.5}
\lim\limits_{K_{a}\rightarrow 0}\Theta(K_{a})= P(K_{a}).
\end{equation}
In general, $\Theta(K_{a})$   may lack coincidence with the
high-energy limit of the momentum-energy tensor trace (if any):
\begin{equation}\label{GT12.3.5new}
\lim\limits_{a\rightarrow\l_{min}} P(K_{a}).
\end{equation}
At the same time, when we naturally assume that the Static
Spherically-Symmetric Horizon Space-Time with the radius of
several Planck’s units (\ref{GT12.3.4new}) is nothing else but a
micro black hole, then the high-energy limit (\ref{GT12.3.5new})
is existing and the replacement of   $\Theta(K_{a})$  by
 $P(K_{a})$  in r.h.s. of the foregoing equations is possible to
give a hypothetical gravitational equation   for the event horizon
micro black hole. But a question arises, for which values of the
parameter $a$ (\ref{GT12.3.4new}) (or $K_{a}$) this is valid and
what is a minimal value of the parameter $\gamma=\gamma(a)$ in
this case.
\\In all the cases under study, 2.1) and   2.2), the deformation parameter
$\alpha_{a}$ (\ref{D1}) ($\lambda_{a}$(\ref{D1.1})) is, within the
constant factor, coincident with the Gaussian curvature $K_{a}$
(respectively $\sqrt{K_{a}}$) that is  in essence continuous in
the low-energy case and discrete in the high-energy case.
\\{\bf In this way the above-mentioned example shows that,
despite the absence of infinitesimal spatial-temporal increments
owing to the existence of $l_{min}$ and the essential
«discreteness» of a theory, this  discreteness at low energies is
not «felt», the theory being actually continuous. The indicated
discreteness is significant only in the case of high (Planck’s)
energies.}
\\
\\In \cite{Padm13} it is shown that
 the Einstein Equation for horizon spaces in the differential
form may be written as a thermodynamic identity (the first
principle of thermodynamics) (\cite{Padm13}, formula (119)):
\begin{equation}\label{GT12}
   \underbrace{\frac{{{\hbar}} cf'(a)}{4\pi}}_{\displaystyle{k_BT}}
    \ \underbrace{\frac{c^3}{G{{\hbar}}}d\left( \frac{1}{ 4} 4\pi a^2 \right)}_{
    \displaystyle{dS}}
  \ \underbrace{-\ \frac{1}{2}\frac{c^4 da}{G}}_{
    \displaystyle{-dE}}
 = \underbrace{P d \left( \frac{4\pi}{ 3}  a^3 \right)  }_{
    \displaystyle{P\, dV}}.
\end{equation}
where, as noted above, $T$ -- temperature of the horizon surface,
$S$ --corresponding entropy, $E$-- internal energy, $V$ -- space
volume.
\\Note that, because of the existing $l_{min}$, practically all quantities in  (\ref{GT12})
(except of $T$) represent GEQ. Apparently, the radius of a sphere
$r=a$, its volume $V$, and entropy represent such quantities:
\begin{equation}\label{GT12.new1}
S=\frac{4\pi a^{2}}{4l^{2}_{P}}=\frac{\pi a^{2}}{l^{2}_{P}},
\end{equation}
within the constant factor $1/4l^{2}_{P}$ equal to a sphere with the radius $a$.
\\ Because of this, there are no infinitesimal increments of these quantities,
i.e.  $da, dV, dS$. And, provided $l_{min}$ is involved, the
Einstein equation for the above-mentioned case in the differential
form (\ref{GT12}) makes no sense and is useless.  If $da$ may be,
purely formerly, replaced by $l_{min}$, then, as the quantity
$l_{min}$ is fixed, it is obvious that «$dS$» and  «$dV$» in
(\ref{GT12}) will be growing as $a$ and $a^{2}$, respectively. And
at low energies,
 i.e. for large values of $a\gg l_{min}$, this naturally
 leads to infinitely large rather than infinitesimal values.
\\ In a similar way it is easily seen that the «Entropic Approach to Gravity»
\cite{Verlinde} in the present formalism is invalid within the
scope of the minimal length theory. In fact, the «main instrument»
in \cite{Verlinde} is a formula for the infinitesimal variation
$dN$ in the bit numbers $N$ on the holographic screen ${\cal S}$
with the radius $R$ and with the surface area $A$
(\cite{Verlinde},formula (4.18)):
\be \label{bitdensity}
dN={c^3\over G\hbar}\, dA=\frac{dA}{l^{2}_P}.
 \ee
 As $N=A/l^{2}_P$, and $A$ represents  GEQ, it is clear that
 $N$ is also GEQ and hence neither $dA$ nor $dN$ makes sense.
\\It is obvious that infinitesimal variations of
the screen surface area $dA$ are possible only in a continuous
theory involving no $l_{min}$.
\\ When $l_{min}\propto l_P$ is involved, the minimal variation
$\triangle A$ is evidently associated with a minimal variation in
the radius $R$ \be \label{bitdensity1} R\rightarrow R\pm l_{min}
\ee  is dependent on $R$ and growing as $R$ for $R\gg l_{min}$
 \be \label{bitdensity2}
\triangle_{\pm} A(R)=(A(R\pm l_{min})-A(R))\propto
(\frac{\pm2R}{l_{min}}+1)=\pm2N_{R}+1,
 \ee
where $N_{R}=R/l_{min}$, as indicated above.
\\ But, as noted above, a minimal increment of the radius $R$ equal to
$|\Delta_{min}R|=l_{min}\propto l_P$  corresponds only to the case
of maximal (Planck’s) energies or, what is the same,  to the
parameter $\gamma = 1$  in formula (\ref{D1.1new}). However, in
\cite{Verlinde} the considered low energies are far from the
Planck energies and hence in this case in (\ref{D1.1new})
$\gamma<1$, (\ref{bitdensity1}), and (\ref{bitdensity2}) are
respectively replaced by \be \label{bitdensity1.1} R\rightarrow
R\pm N_{R}^{1-\gamma}l_{min}
 \ee
 and
 \be \label{bitdensity2.1}
\triangle_{\pm} A(R)=(A(R\pm N_{R}^{1-\gamma}l_{min})-A(R))\propto
\pm N_{R}^{2-\gamma}+N_{R}^{2-2\gamma}=N_{R}^{2-2\gamma}(\pm
N_{R}^{\gamma}+1).
 \ee
An increase of r.h.s in(\ref{bitdensity2.1}) with the growth of
$R$ (or identically of $N_{R}$) for $R\gg l_{min}$  is obvious.
\\So, if $l_{min}$ is involved, formula
(4.18) from \cite{Verlinde} makes no sense similar to other formulae
derived on its basis (4.19),(4.20),(4.22),(5.32)--(5.34), … in
\cite{Verlinde} and similar to the derivation method for
Einstein’s equations proposed in this work.
\\Proceeding from the principal parameters of this work $\alpha_{l} (or \lambda_{l})$,
the fact is obvious and is supported by the formula
 (\ref{Beken1}) given in this paper, meaning that
 \begin{equation}\label{Beken2}
\alpha^{-1}_{R}\sim A,
\end{equation}
i.e. small variations of  $\alpha_{R}$ (low energies) result in
large variations of $\alpha^{-1}_{R}$, as indicated by formula
(\ref{bitdensity2}).
\\In fact, we have a {\bf no-go theorems}.
\\ The last statements concerning $dS,dN$ may be explicitly
interpreted using the language of a quantum information theory as
follows:
\\{\bf due to the existence of the minimal length $l_{min}$, the minimal area
$l^{2}_{min}$ and volume $l^{3}_{min}$ are also involved, and that
means «quantization» of the areas and volumes}. As, up to the
known constants, the «bit number» $N$ from (\ref{bitdensity}) and
the entropy   $S$ from (\ref{GT12.new1}) are nothing else but
 \be \label{bitdensity3.1}
S=\frac{A}{4l^{2}_{min}},N=\frac{A}{l^{2}_{min}},
 \ee
it is obvious that there is a «minimal measure» for the «amount of data»
that may be referred to as «one bit» (or «one qubit»).
\\ The statement that there is no such quantity as $dN$
(and respectively $dS$) is equivalent to claiming the absence of
$0.25$ bit, $0.001$ bit, and so on.
\\ This inference completely conforms to the Hooft-Susskind Holographic
Principle (HP) \cite{Hooft1}--\cite{Bou1} that includes two main statements:
\\
\\(a)All information contained in a particular spatial domain
is concentrated at the boundary of this domain.
\\
(b)A theory for the boundary of the spatial domain under study
should contain maximally one degree of freedom per Planck’s area
$l^{2}_{P}$.
\\
\\ In fact (but not explicitly) HP implicates the existence of
$l_{min}=l_{P}$. The existence of $l_{min}\propto l_{P}$  totally
conforms to HP, providing its generalization. Specifically,
without the loss of generality, $l^{2}_{P}$ in point (b) may be
replaced by  $l^{2}_{min}$.
\\
\\ So, the principal inference of this work is as follows:
\\ provided the minimal length $l_{min}$ is involved, its existence must
be taken into consideration not only at high but also at low
energies, both in a quantum theory and in gravity. This becomes
apparent by rejection of the infinitesimal quantities associated
with the spatial-temporal variations $dx_{\mu},...$. In other
words, with the involvement of $l_{min}$, the General Relativity
(GR) must be replaced by a (still unframed) minimal-length
gravitation theory that may be denoted as $Grav^{l_{min}}$. In
their results GR and $Grav^{l_{min}}$  should be very close but,
as regards their mathematical apparatus (instruments), these
theories  are absolutely different.
\\ Besides, $Grav^{l_{min}}$ should offer a rather natural
transition from high to low energies
\begin{equation}\label{Concl-2}
[N_{L}\approx 1]\rightarrow [N_{L}\gg 1]
\end{equation}
and vice versa
\begin{equation}\label{Concl-2.1}
 [N_{L}\gg 1]\rightarrow [N_{L}\approx 1],
\end{equation}
where $N_{L}$ -- integer from formula (\ref{Introd 2.4})
determining the characteristics scale of the lengths $L$ (energies
$E\sim 1/L\propto 1/N_{L})$.

\begin{center}
{\bf Conflict of Interests}
\end{center}
The author declares that there is no conflict of interests
regarding the publication of this article.

\end{document}